# Revolutionizing Networking Paradigms: A Comprehensive Exploration of Information-Centric Networking (ICN), Content-Centric Networking (CCNx) and Named Data Networking (NDN)

Kamorudeen Amuda*, Wakili Almustapha[†], Binkam Deepak[‡], Ciana Hoggard[§], Pranay Tiruveedula[¶]
Computer and Information Sciences Department
Towson University
Baltimore, USA
Email: {*kamuda1, [†]awakili1, [‡]dbinkam1, [§]chogga1, [¶]ptiruve1}@students.towson.edu

*Abstract*—The evolution of networking paradigms has led to the emergence of Information-Centric Networking (ICN), Content-Centric Networking (CCNx), and Named Data Networking (NDN). These innovative architectures move away from traditional host-centric models to focus on content-oriented approaches. This paper offers a succinct understanding and in-depth exploration of these revolutionary networking frameworks. It explores into their foundational concepts, technical principles, and the latest developments. Also, the paper examines the current standards that govern these architectures and discusses their practical applications in the real world. By providing a comprehensive overview, this study aims to highlight how ICN, CCNx, and NDN are reshaping the landscape of digital communications, emphasizing efficiency, security, and scalability in data handling and distribution across networks.

## I. INTRODUCTION

The rise of the Internet of Things (IoT) has led to unprecedented growth in data generation. This surge has created a pressing demand for seamless communication between producer and consumer nodes across a wide range of network environments, such as Wireless Sensor Networks (WSNs) and mobile environments. However, traditional Internet architectures reliant on host-centric address-based communication face several obstacles, such as privacy concerns, security risks, routing inconsistencies, and mobility. The increasing flow of mobile data highlights the importance of mobility in networking, prompting the demand for innovative solutions to address its challenges while ensuring efficient data delivery.

The traditional network model heavily relies on IP addresses to establish connections between devices on the Internet. Location-oriented data communication often results in delays and excessive bandwidth consumption, leading to increased power, energy, and resource usage. In addition, the architecture of the IP-based Internet is impacted by issues such as network congestion and the proliferation of identical content in multiple distant locations. Given its significant resource and energy demands, these challenges highlight the shortcomings of a host-centric Internet architecture in distributing data efficiently. Furthermore, the emergence of novel applications that feature heterogeneous data further complicates the distribution of data through constrained network channels. The transition from traditional networking models to Information-Centric Networking (ICN) represents a significant paradigm shift in how data is handled across networks. Traditional models such as the OSI and TCP/IP frameworks focus on data transmission between specific hosts, utilizing IP addresses to route packets through various network paths. These models are primarily host-centric, which can lead to inefficiencies in modern networking scenarios that demand more dynamic content delivery methods. The following frameworks address the limitations of traditional models by focusing on the efficient and secure delivery of content directly to the network [1].

**Information-Centric Networking (ICN)** introduces a novel approach by making content, rather than hosts, the central element of network architecture. In ICN, data is accessed and disseminated based on its name, not its location. This shift is embodied in two prominent ICN architectures: Content-Centric Networking (CCNx) and Named Data Networking (NDN), which both prioritize content retrieval based on names across the network [2].

**Content-Centric Networking (CCNx)** focuses on network protocols that use Interest and Content Objects to optimize how requests and content are managed. The core concept involves a consumer sending out an Interest packet specifying the content by name, and the network responding with the named Content Object. This method reduces the need for a direct link between the consumer and the content's original host, instead allowing the content to be fetched from the nearest node that has it cached.

**Named Data Networking (NDN)** shares foundational principles with CCNx but emphasizes the infrastructure's role in caching and delivering content directly from within the network nodes. This not only simplifies content delivery but also enhances security and efficiency. NDN enables data packets to be securely signed and named, ensuring content authenticity and integrity directly through network protocols.

The importance of transitioning to Content-Centric Networking stems from its numerous advantages over traditional host-centric models.

1) **Efficiency:** CCN supports the efficient use of network resources by enabling ubiquitous in-network caching. This caching reduces redundancy, decreases reliance on original servers, and minimizes data transmission costs.
2) **Scalability:** By decoupling content from its location, CCN naturally supports scalability. Networks can handle increased traffic by simply catching popular content closer to the user, reducing the distance data must travel and balancing loads more effectively across the network.
3) **Security:** In CCN, security is inherently content-centric. Data packets include cryptography signatures that ensure content integrity and authenticity regardless of where the content is stored or how it is transmitted. This built-in security feature is vital in a digital landscape where trust and security are paramount.
4) **Robustness:** CCN enhances network robustness by facilitating efficient data delivery even in challenging conditions such as mobile environments and intermittent connectivity. This is achieved through native support for multi-cast and efficient data dissemination mechanisms.

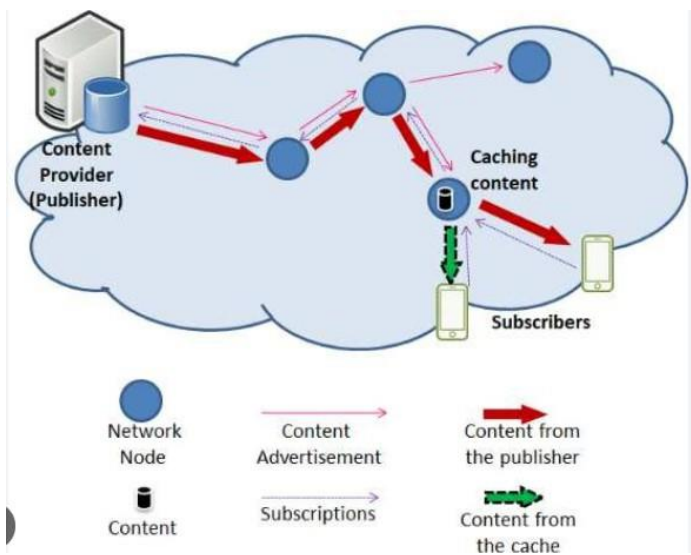

Fig. 1. Information-centric networking model [3].

The Information-Centric Networking (ICN) model primarily focuses on enhancing the efficiency of content delivery across the network. It comprises several key components and interactions:

- **Content Provider (Publisher):** The origin of the content, is typically represented as a server.
- **Network Cloud:** Depicts the network infrastructure which facilitates the flow of content and requests.
- **Caching Content:** Indicates that content can be cached within the network to improve data retrieval speed and reduce bandwidth usage. This is symbolically represented by a secure lock in the network diagram.
- **Subscribers:** End users or devices that request and consume the content, shown as mobile devices in the diagram.
- **Content Flows:**
  - Red arrows represent the distribution of content from the publisher to the network and from the network to subscribers.
  - Green arrow illustrates the efficient delivery of cached content directly from the network to subscribers, bypassing the original content provider.
  - Pink arrows denote the advertisement of available content and the subscription requests flowing through the network.

As the Internet evolves towards more content-driven services, the need for efficient, secure, and scalable networking solutions becomes crucial. CCN offers a forward-thinking framework that addresses these needs by rethinking traditional IP-based approaches and focusing on how data are consumed rather than how it is delivered. This transition not only aligns with current digital consumption patterns but also sets the stage for future innovations in network technology [4].

## II. KEY CONCEPTS AND PRINCIPLES

ICN introduces a paradigm in which data are addressed and retrieved by their content name rather than by the host's address, fundamentally differentiating it from traditional IP-based networking. This approach allows for more efficient data retrieval and distribution, especially in environments where content availability and delivery speed are critical. Key principles of ICN include content-based naming and security, in-network caching, and name-based content discovery and delivery. These principles collectively improve network efficiency by reducing redundancy and improving security at the data level.

### A. ICN Architectures

The various architectures underpin the ICN framework, with Named Data Networking (NDN) and Content-Centric Networking (CCNx) being prominent. These architectures share a mechanism where data are retrieved by issuing interest packets that traverse the network until they locate the desired content. Data packets then retrace the path of the interest to deliver the content to the requester. This method efficiently utilizes network resources by caching data along the path, thus speeding up data delivery and reducing network load [5].

### B. Comparison with Traditional Networking

Traditional networking, based on the TCP/IP model, routes communications based on fixed IP addresses of hosts, leading to inefficiencies in dynamic and mobile environments. In contrast, ICN does not tie data retrieval to host locations, but to the content itself, which supports mobility and reduces the overhead associated with managing IP addresses and routes. This makes ICN particularly suitable for modern applications such as streaming media, content delivery networks, and mobile applications where users frequently change locations [6].

*C. Content Naming, Routing, and Caching*

ICN uses a unique naming scheme for content, which can be hierarchical or flat, providing flexibility and efficiency in content retrieval. Routing in ICN is performed based on these names rather than network addresses, allowing for dynamic path selection based on network conditions and content location. Caching is integral to ICN, and content is cached at multiple points in the network to facilitate rapid retrieval. This distributed caching mechanism not only improves the speed of data access, but also significantly reduces the bandwidth required for content delivery, as popular content is stored closer to potential requesters [6].

In summary, ICN represents a significant shift from traditional host-centric network architectures towards a more flexible, efficient, and secure content-centric networking approach. This shift addresses many of the challenges of modern networking, including the need for efficient content distribution, improved security, and better support for mobile and distributed applications. As the volume of content delivered over the internet continues to grow, ICN offers an effective and efficient solution that aligns well with the future needs of global data communication.

## III. TECHNICAL UNDERPINNINGS

The technical underpinnings of Named Data Networking (NDN), within the broader context of Information-Centric Networking (ICN), reflect a significant paradigm on how data are addressed, routed, and retrieved over networks. NDN, along with its related protocols, such as CCNx (Content-Centric Networking), embodies a comprehensive rethinking of traditional network architecture, focusing heavily on the content itself rather than the location of where the content is stored. This synthesis explores NDN's key components:

*A. Addressing and Naming Schemes*

The fundamental aspect of NDN is its innovative approach to addressing and naming, which is markedly different from traditional IP-based systems. Unlike IP addresses, which are tied to device locations, NDN uses human-readable, hierarchical names to address content. This naming scheme supports efficient data retrieval and caching mechanisms, using structured names that enhance network efficiency. For instance, hierarchical names facilitate aggregation, significantly reducing the size of routing tables and simplifying the management of network resources.

Moreover, NDN supports various naming schemes—hierarchical, flat, attribute-value, and hybrid—each catering to specific network needs and scenarios.

*1) Hierarchical naming:* Hierarchical naming is particularly advantageous for aggregating data under common prefixes, thus optimizing the forwarding process.

*2) Flat naming:* Flat naming, often derived through hashing mechanisms, provides unique and secure identifiers suitable for dynamic environments.

*3) Attribute-value and hybrid schemes:* it offer a blend of scalability and flexibility, ensuring robust support for diverse applications ranging from multimedia streaming to large-scale content distribution networks [7].

*B. Forwarding and Routing Mechanisms*

NDN's forwarding and routing mechanisms are fundamentally centered around content names instead of endpoint identifiers. Interest packets, which are used to request content, navigate the network based on content names embedded within them. These packets are processed by routers that maintain a Forwarding Information Base (FIB), which directs the requests towards potential data sources. This approach eliminates the reliance on a single data source, enhancing data availability and network resilience.

The architecture also includes a Pending Interest Table (PIT) that tracks all outgoing interest requests and incoming data packets, ensuring that data can be efficiently retrieved and forwarded to the requester. Additionally, the Content Store (CS) within each router caches the data passing through, which can significantly reduce retrieval times and reduce network congestion during peak times. This integrated caching mechanism is crucial to reducing latency and improving the overall efficiency of content delivery across the network [8].

*C. Content Retrieval and Delivery Process*

The content retrieval process in NDN is consumer-driven; consumers send out Interest packets that specify the content they wish to access. These packets traverse the network, retrieving data from the nearest node that can fulfill the request, which might be an intermediate router that has cached the content, not necessarily the original source. This in-network caching capability is a distinctive feature of NDN, enabling quicker access to content and reducing the amount of data that needs to traverse the entire network.

NDN also employs cooperative caching strategies, such as neighbourhood cooperative caching, where multiple network nodes collaborate to determine the most effective locations for caching content. This cooperation helps optimize the use of available cache resources, improving network throughput, and reducing content access latency. Such strategies are particularly effective in diverse network environments where content demand can be high and varied between different network segments [8].

Through these mechanisms, NDN addresses the inefficiencies of traditional IP networks by focusing on the content itself rather than the locations of the content servers. This approach not only enhances data security, given that content can be authenticated independently of its location, but also improves scalability and network management. As a result, NDN is particularly well-suited to modern digital ecosystems, which are characterized by vast amounts of data and a high demand for efficient, reliable content delivery.

## D. Recent Developments and Standards

Recent advances in ICN, CCNx, and NDN emphasize the importance of efficient content delivery protocols. ICN, particularly, has seen the integration of blockchain technology to enhance content security and integrity. This integration is beneficial in managing efficient data retrieval and providing robust data security, which are essential for supporting security in IoT environments [7]. The standardization of ICN-related protocols has been critical in fostering interoperability and functionality between different network implementations. RFC 8569 and RFC 8793 are pivotal in this regard. RFC 8569 outlines the core principles of CCNx, providing a structured approach to content naming and caching that enhances data retrieval processes across networks. RFC 8793 focuses on load-balancing strategies in NDN, ensuring efficient data handling and distribution without overloading network nodes. These standards are essential for developers and network engineers to build interoperable systems that adhere to established guidelines and protocols [9].

Moreover, the ongoing development and refinement of these protocols are supported by additional documentation and RFC proposals that address specific aspects of ICN operations, such as security enhancements, routing efficiency, and integration with traditional IP networks, ensuring that these new paradigms can coexist with and improve existing network architectures [4]

*1) Overview of RFCs and Standards:* RFC 8793 is essential in providing standardized terminology for understanding Information-Centric Networking (ICN), Content-Centric Networking (CCNx) and Named Data Networking (NDN). It defines crucial terms like 'Interest Packets,' 'Data Packets,' and 'Forwarding Strategies.' RFC 8793 establishes a standard vocabulary that allows for clear and consistent discussions among ICN researchers. This standardization is essential for advancing ICN studies and implementations, which improve data retrieval efficiency through caching and routing mechanisms that focus on data names instead of host addresses.

RFC 8569 outlines the principles of CCNx within the broader framework of ICN. The document provides a protocol that operates through two types of messages: Interests and Content Objects. Interest messages are sent by consumers to request content by name, and Content Objects are responses to these requests containing the requested data. The document also includes relevant consumer, publisher, and forwarder behaviors. The document emphasizes that, unlike traditional IP-based systems, CCNx employs hierarchical names to manage and route requests and match responses, rather than endpoint addresses. This name-based system involves the forwarding of requests for content directly associated with these names, rather than relying on the location of the content. This allows for efficient data retrieval and distribution since the content can be cached strategically across the network, making it readily accessible near the requestors and reducing redundancy.

ICN and its derivatives, such as CCNx and NDN, have been the focus of ongoing development aimed at enhancing the efficiency and security of content delivery across networks. One notable advancement is the incorporation of blockchain technology into ICN frameworks, which strengthens security measures and integrity checks, making these networks particularly adept at handling secure transactions and data exchanges in environments such as the Internet of Things (IoT) and other digital ecosystems that require robust data integrity mechanisms [10].

*2) Integration with Emerging Technologies:* ICN's integration with cutting-edge technologies such as 5G and Multi-Access Edge Computing (MEC) illustrates its adaptability and relevance in modern network settings. Using the content-centric nature of ICN within 5G networks, operators can significantly reduce latency and improve data throughput, which is critical for real-time applications like streaming, gaming, and autonomous vehicle communications. Similarly, MEC benefits from ICN's efficient data distribution methods to bring computational resources closer to the data source, enhancing the performance of edge-based applications and services.

These integrations are supported by standardization efforts from major telecommunications and standards bodies, such as the European Telecommunications Standards Institute (ETSI) and the Third Generation Partnership Project (3GPP). These organizations play a crucial role in defining the technical specifications and guidelines that ensure that these technologies work seamlessly within broader telecommunication ecosystems and support the deployment of next-generation network technologies [7].

## E. Case Studies and Applications

The evolution of networking paradigms towards more efficient, secure, and scalable systems has driven the development of Information-Centric Networking (ICN), including its two significant branches: Content-Centric Networking (CCNx) and Named Data Networking (NDN). These paradigms shift the focus from where the content is (i.e., address-based retrieval) to what the content is (i.e., content-based retrieval). Below are detailed case studies across different sectors that demonstrate the practical implementations and benefits of these innovative networking paradigms.

Real-World Deployment: The following sections explore various real-world deployments based on academic research and practical implementations outlined in the recent literature.

*1) Smart Home Networks:* ICN's deployment in smart home environments showcases its advantages over traditional IP-based systems, particularly in terms of security, mobility, and multicasting. The discussion of how ICN simplifies connectivity and enhances security in smart homes highlights the critical application of ICN to meet the complex demands of modern connected homes [6].

*2) Internet of Things (IoT):* The Internet of Things (IoT) represents a prime area for ICN application, addressing challenges such as device heterogeneity and scalability. The research highlights ICN's ability to reduce network complexity and enhance efficiency by enabling devices to communicate based on content names rather than network addresses. This capability is particularly advantageous in scenarios where device mobility or frequent changes in network endpoints occur, with ICN caching mechanisms further reducing latency and congestion [6].

*3) Vehicular Networks (VANETs):* In Vehicular Ad hoc Networks (VANETs), ICN enhances safety and traffic management by facilitating efficient and secure data exchange between vehicles. The ability to cache data directly within the network allows for rapid dissemination of critical information, improving traffic safety and management. This deployment underlines ICN's utility in urban settings where real-time communication between vehicles is essential for effective traffic and safety management.

*4) Content Distribution Networks (CDNs):* ICN significantly impacts CDNs by optimizing content delivery, reducing redundancy, and enhancing bandwidth usage. Integrating ICN with CDNs improves the handling of dynamic content demands, especially in multimedia streaming services. The efficient caching and replication strategies inherent in ICN reduce latency and server load, which enhances user experience and decreases operational costs.

*5) Emergency Response Networks:* The decentralized data-sharing capabilities of the ICN are particularly valuable in emergency response scenarios. The resilience of ICN ensures reliable communication even when traditional infrastructures fail. During disasters, the ICN maintains essential communications and coordination among emergency responders, demonstrating its potential to function effectively under adverse conditions.

*6) Healthcare Systems:* ICN also plays a pivotal role in healthcare systems, managing sensitive data securely and efficiently. Real-time patient monitoring and secure medical record access are facilitated by ICN, enhancing medical response and routine health monitoring's efficacy and responsiveness. This deployment highlights ICN's capability to address critical healthcare delivery aspects, ensuring data accessibility and security.

In a nutshell, ICN's real-world deployments across various domains demonstrate its capacity to meet the demands of the modern network environment. From smart homes and IoT to VANETs, CDNs, emergency response, and healthcare, ICN provides substantial improvements over traditional network architectures. These case studies not only underscore the versatility and scalability of ICN but also affirm its effectiveness and efficiency, positioning it as a foundational technology for the future of networking.

## IV. CONCLUSION

The development of Information-Centric Networking (ICN) and its derivatives, Content-Centric Networking (CCNx) and Named Data Networking (NDN) marks a significant advancement in networking paradigms, moving from traditional host-centric methods to content-centric approaches. This change is vital because it significantly enhances the efficiency, security, and scalability of networks by focusing on content rather than location. Such a focus is especially relevant to the demands of the modern digital age. ICN improves data delivery and streamlines network architectures, which is beneficial for emerging technologies like 5G and IoT. The implementation of ICN also addresses several prevalent challenges in network infrastructures, including scalability and security vulnerabilities. By centralizing content, ICN simplifies the complexity traditionally involved in network management, making it a powerful tool for improving network performance and security across various platforms.

## REFERENCES

[1] M. S. M. Shah, Y.-B. Leau, Z. Yan, and M. Anbar, "Hierarchical naming scheme in named data networking for internet of things: A review and future security challenges," *IEEE access*, vol. 10, pp. 19958–19970, 2022.

[2] B. Wissingh, C. Wood, A. Afanasyev, L. Zhang, D. Oran, and C. Tschudin, "Rfc 8793: Information-centric networking (icn): Content-centric networking (ccnx) and named data networking (ndn) terminology," 2020.

[3] Y. Fazea and F. Mohammed, "Software defined networking based information centric networking: An overview of approaches and challenges," in *2021 International Congress of Advanced Technology and Engineering (ICOTEN)*, pp. 1–8, IEEE, 2021.

[4] M. Mosko, I. Solis, and C. Wood, "Rfc 8569: Content-centric networking (ccnx) semantics," 2019.

[5] A. Afanasyev, I. Moiseenko, L. Zhang, *et al.*, "ndnsim: Ndn simulator for ns-3," *University of California, Los Angeles, Tech. Rep*, vol. 4, pp. 1–7, 2012.

[6] M. Amadeo, C. Campolo, J. Quevedo, D. Corujo, A. Molinaro, A. Iera, R. L. Aguiar, and A. V. Vasilakos, "Information-centric networking for the internet of things: challenges and opportunities," *IEEE Network*, vol. 30, no. 2, pp. 92–100, 2016.

[7] O. Serhane, K. Yahyaoui, B. Nour, and H. Moungla, "A survey of icn content naming and in-network caching in 5g and beyond networks," *IEEE Internet of Things Journal*, vol. 8, no. 6, pp. 4081–4104, 2020.

[8] P. Chaudhary, N. Hubballi, and S. G. Kulkarni, "Ncache: neighborhood cooperative caching in named data networking," in *2022 5th International Conference on Hot Information-Centric Networking (HotICN)*, pp. 36–41, IEEE, 2022.

[9] S. Mastorakis, D. R. Oran, J. Gibson, I. Moiseenko, and R. Droms, "Information-Centric Networking (ICN) Ping Protocol Specification." RFC 9508, Mar. 2024.

[10] M. S. M. Shah, Y.-B. Leau, M. Anbar, and A. A. Bin-Salem, "Security and integrity attacks in named data networking: a survey," *IEEE Access*, vol. 11, pp. 7984–8004, 2023.